\documentclass[aps,prl,reprint,superscriptaddress,nofootinbib]{revtex4-2}

\usepackage[utf8]{inputenc}
\usepackage{standalone}

\usepackage{xspace}

\usepackage{makecell}
\usepackage{orcidlink}

\usepackage{amsmath}
\usepackage{newtxtext,newtxmath}
\usepackage{siunitx}
\usepackage[version=4]{mhchem}

\usepackage{color}
\usepackage{braket}

\newcommand{\Nd}{$^{150}$Nd\xspace}
\newcommand{\Sm}{$^{154}$Sm\xspace}
\newcommand{\higs}{HI$\gamma$S\xspace}
\newcommand{\gthree}{$\gamma^3$}
\newcommand{\gray}{$\gamma$-ray\xspace}
\newcommand{\LaBr}{LaBr\textsubscript{3}}
\DeclareSIUnit\barn{b}
\DeclareSIUnit\nuclearmagnetron{\text{\ensuremath{\mu_\text{N}}}}
\DeclareSIUnit\nuclearmagnetronsquared{\text{\ensuremath{\mu_\text{N}^2}}}

\makeatletter
\newcommand\footnoteref[1]{\protected@xdef\@thefnmark{\ref{#1}}\@footnotemark}
\makeatother

\definecolor{prlblue}{HTML}{2E3092}
\usepackage{hyperref}
\hypersetup{unicode,
    pdfdisplaydoctitle,
    breaklinks=true,
    colorlinks=true,
    allcolors=prlblue,
    pdftitle={Deviations from the Porter-Thomas distribution due to non-statistical γ decay below the ¹⁵⁰Nd neutron separation threshold},
    pdfauthor={}
}

\begin{document}

\title{Deviations from the Porter-Thomas Distribution due to Nonstatistical \(\gamma\) Decay \\ below the \Nd{} Neutron Separation Threshold}

\newcommand{\ikp}{\href{https://ror.org/05n911h24}{Technische Universität Darmstadt},
	Department of Physics,
	Institute for Nuclear Physics,
	64289 Darmstadt,
	Germany}
\newcommand{\wnsl}{Wright Nuclear Structure Laboratory,
	Yale University,
	New Haven,
	CT 06520,
	USA}
\newcommand{\notredame}{University of Notre Dame,
	Notre Dame,
	IN 46556,
	USA}
\newcommand{\delhi}{Department of Physics and Astrophysics,
	University of Delhi,
	Delhi 110007,
	India}
\newcommand{\dopakentucky}{Department of Physics and Astronomy,
	University of Kentucky,
	Lexington,
	KY 40506-0055,
	USA}
\newcommand{\dockentucky}{Department of Chemistry,
	University of Kentucky,
	Lexington,
	KY 40506-0055,
	USA}
\newcommand{\duke}{Department of Physics,
	\href{https://ror.org/00py81415}{Duke University},
	Durham,
	NC 27708-0308,
	USA}
\newcommand{\tunl}{\href{https://ror.org/00ax83b61}{Triangle Universities Nuclear Laboratory},
	Durham,
	NC 27708-0308,
	USA}
\newcommand{\uncch}{Department of Physics and Astronomy,
	\href{https://ror.org/0130frc33}{University of North Carolina at Chapel Hill},
	Chapel Hill,
	NC 27599,
	USA}
\newcommand{\gsi}{\href{https://ror.org/02k8cbn47}{GSI Helmholtzzentrum für Schwerionenforschung GmbH},
	64291 Darmstadt,
	Germany}
\newcommand{\uws}{School of Computing,
	Engineering, and Physical Sciences,
	\href{https://ror.org/04w3d2v20}{University of the West of Scotland},
	Paisley PA1 2BE,
	United Kingdom}
\newcommand{\supa}{\href{https://ror.org/026efkc57}{The Scottish Universities Physics Alliance},
	Glasgow G12 8QQ,
	United Kingdom}
\newcommand{\nscl}{National Superconducting Cyclotron Laboratory,
	\href{https://ror.org/05hs6h993}{Michigan State University},
	East Lansing,
	MI 48824,
	USA}
\newcommand{\fsu}{Department of Physics,
	Florida State University,
	Tallahassee,
	FL 32306,
	USA}
\newcommand{\koeln}{Institut für Kernphysik,
	\href{https://ror.org/00rcxh774}{Universität zu Köln},
	50937 Köln,
	Germany}
\newcommand{\usg}{\href{https://ror.org/00n3w3b69}{University of Strathclyde},
	Glasgow G4 0NG,
	United Kingdom}
\newcommand{\ess}{European Spallation Source ERIC,
	224 84, Lund,
	Sweden}
\newcommand{\kuleuven}{KU Leuven,
	Instituut voor Kern- en Stralingsfysica,
  3001 Leuven,
  Belgium}
\newcommand{\lmu}{Ludwig-Maximilians-Universität München,
	Universitäts-Sternwarte,
	81679 Munich,
	Germany}
\newcommand{\mpigarching}{Max-Planck Institute for Extraterrestrial Physics,
	85748 Garching,
	Germany}
\newcommand{\vysus}{Vysus Group Sweden AB,
  214 21 Malmö,
  Sweden}

\author{O.~\surname{Papst}\,\orcidlink{0000-0002-1037-4183}}
\email{opapst@ikp.tu-darmstadt.de}
\affiliation{\ikp}

\author{J.~\surname{Isaak}\,\orcidlink{0000-0002-4735-8320}}
\email{jisaak@ikp.tu-darmstadt.de}
\affiliation{\ikp}

\author{V.~\surname{Werner}\,\orcidlink{0000-0003-4001-0150}}
\affiliation{\ikp}

\author{D.~\surname{Savran}\,\orcidlink{0000-0002-2685-5600}}
\affiliation{\gsi}

\author{N.~\surname{Pietralla}\,\orcidlink{0000-0002-4797-3032}}
\affiliation{\ikp}

\author{G.~\surname{Battaglia}\,\orcidlink{0000-0003-1974-8521}}
\affiliation{\usg}

\author{T.~\surname{Beck}\,\orcidlink{0000-0002-5395-9421}}
\altaffiliation[Present address: ]{\kuleuven}
\affiliation{\ikp}

\author{M.~\surname{Beuschlein}\,\orcidlink{0009-0001-3183-7503}
}
\affiliation{\ikp}

\author{S.~W.~\surname{Finch}\,\orcidlink{0000-0003-2178-9402}}
\affiliation{\duke}
\affiliation{\tunl}

\author{U.~\surname{Friman-Gayer}\,\orcidlink{0000-0003-2590-5052}}
\altaffiliation[Present address: ]{\vysus}
\affiliation{\ikp}

\author{K.~E.~\surname{Ide}\,\orcidlink{0000-0003-2405-329X}}
\affiliation{\ikp}

\author{R.~V.~F.~\surname{Janssens}\,\orcidlink{0000-0001-7095-1715}}
\affiliation{\tunl}
\affiliation{\uncch}

\author{M.~D.~\surname{Jones}}
\affiliation{\tunl}
\affiliation{\uncch}

\author{J.~\surname{Kleemann}\,\orcidlink{0000-0003-2596-3762}}
\affiliation{\ikp}

\author{B.~\surname{L{\"o}her}\,\orcidlink{0000-0002-9010-2558}}
\affiliation{\gsi}

\author{M.~\surname{Scheck}\,\orcidlink{0000-0002-9624-3909}}
\affiliation{\uws}
\affiliation{\supa}

\author{M.~\surname{Spieker}\,\orcidlink{0000-0002-7214-7656}}
\altaffiliation[Present address: ]{\fsu}
\affiliation{\nscl}

\author{W.~\surname{Tornow}\,\orcidlink{0000-0003-4031-6926}}
\affiliation{\duke}
\affiliation{\tunl}

\author{R.~\surname{Zidarova}\,\orcidlink{0009-0008-1677-9262}}
\affiliation{\ikp}

\author{A.~\surname{Zilges}\,\orcidlink{0000-0002-9328-799X}}
\affiliation{\koeln}

 \date{\today}

\begin{abstract}
    We introduce a new method for the study of fluctuations of partial transition widths
    based on nuclear resonance fluorescence experiments
    with quasimonochromatic linearly polarized photon beams
    below particle separation thresholds.
    It is based on the average branching of decays of \(J=1\) states of an even-even nucleus to the \(2^+_1\) state
    in comparison to the ground state.
    Between \num{5} and \qty{7}{\MeV},
    a constant average branching ratio for $\gamma$ decays from $1^-$ states of
    \num{0.490(16)} is observed
    for the nuclide \Nd.
    Assuming \(\chi^2\)-distributed partial transition widths,
    this average branching ratio is related
    to a degree of freedom of \(\nu = \num{1.93(12)}\),
    rejecting the validity of the Porter-Thomas distribution,
    requiring \(\nu=1\). 
    The observed deviation can be explained by nonstatistical effects in the \(\gamma\)-decay behavior with contributions in the range of \qty{9.4(10)}{\percent} up to \qty{94(10)}{\percent}.
\end{abstract}
 
\maketitle

\textit{Introduction---}Random Matrix Theory (RMT)
provides a comprehensive framework
for the description of complex, chaotic quantum systems~\cite{Weidenmueller2009,Guhr1998}.
It is  exploited across various domains of physics 
ranging from the statistical treatment of nuclear reactions~\cite{Mitchell2010} to recent applications 
in quantum computing~\cite{Boixo2018, Bouland2019, Bulchandani2024,Claeys2025}. 
Developed by Wigner~\cite{Wigner1955} and Dyson~\cite{Dyson1962_I,Dyson1962_II,Dyson1962_III},
RMT considers Hamiltonian matrices with randomly distributed matrix elements. 
An ensemble of such matrices allows for the extraction of statistical properties such as the distribution of level spacings and partial quantum transition widths.
Of particular interest are Gaussian-distributed matrix elements,
resulting in \(\chi^2\)-distributed partial transition widths \(\Gamma_i\)
(see Supplemental Material~\cite{SupplementAPS}\nocite{Goriely2019db}\nocite{Bartholomew1973}\nocite{Alaga1955}\nocite{Guttormsen2021}).
The symmetries conserved by the interaction determine
the degree of freedom \(\nu\) of the \(\chi^2\) distribution.
For integer-valued angular momentum and good time-reversal symmetry,
an orthogonal Hamiltonian with a Gaussian orthogonal ensemble (GOE) is obtained,
resulting in a \(\chi^2\) distribution with one degree of freedom (\(\nu=1\)),
which is commonly referred to as the Porter-Thomas (PT) distribution~\cite{PorterThomas1956}.

A multitude of applications requires information on 
statistical properties of nuclear reaction rates. 
They are needed for quantitative estimates of astrophysical nucleosynthesis processes~\cite{Smith2023, Rochman2025}, for the simulation of nuclear fuel cycles for reactor design~\cite{Alhassan2016} and the transmutation of nuclear waste~\cite{Salvatores2011}, or for nuclear safeguards~\cite{Kolos2022}.
Correctly predicting nuclear reaction rates requires nuclear structure information, including nuclear level densities, and fluctuation properties of partial transition widths. 
Nuclear reaction models that are based on the Hauser-Feshbach formalism~\cite{HauserFeshbach1952} and corresponding computer codes, such as \textsc{Talys-2.0}~\cite{Koning2023}, 
rely critically on the validity of approximating statistical quantum properties by RMT and the PT distribution. 
Therefore, it is essential to benchmark statistical nuclear properties with accurate and reliable nuclear data.

One important aspect in the practical application of Hauser-Feshbach codes is the fluctuation property of partial transition widths.
Neutron-induced nuclear reaction cross sections provided a vast amount of precision data.
Experimentally, the applicability of PT fluctuations
has been extensively studied in thermal neutron capture experiments~\cite{
Haq1982,Adams1998,Shriner2000,Koehler2004,Koehler2007,Koehler2010,Weidenmueller2010,Koehler2011,Celardo2011,Koehler2012,Volya2015,Bogomolny2017,Hagino2021}.
An analysis of the nuclear data ensemble (NDE) of neutron resonances~\cite{Haq1982}
to validate the PT distribution
finds good agreement
between GOE predictions and experimental data.
However, recent studies and thorough reanalyses of the NDE revealed significant deviations from PT predictions;
see, e.g., Refs.~\cite{Adams1998,Shriner2000,Koehler2010,Koehler2011}. The observed degrees of freedom of the underlying $\chi^2$ distribution vary between $\nu \approx 0.5$~\cite{Koehler2010} and $\nu = 2.7(6)$~\cite{Koehler2004, Koehler2007, Koehler2012} and deviate significantly from the PT distribution with $\nu = 1$.
Explanations were suggested that involved  coupling to other decay channels such as nonstatistical $\gamma$ decays. These, in turn,  result in a significant modification of partial transition width distributions (see Refs.~\cite{Weidenmueller2010,Volya2015,Bogomolny2017} and references therein).

Another fundamental question in quantum many-body physics is whether chaotic strongly coupled quantum systems are adequately understood, particularly those in which nuclear deformation is expected to preserve $K$ quantum numbers.
The influence of the $K$ quantum numbers on $\gamma$ decay subsequent to thermal and average-resonance neutron capture is a subject of considerable debate~\cite{Rekstad1990, Hansen1991, Barrett1992, Sheline1995, Huseby1997}. 
It is often expected that nuclear levels at excitation energies in the region of neutron resonances are completely mixed and thus exhibit a chaotic quantum structure.
While it is commonly assumed that the $\gamma$-decay behavior is well described within a statistical ansatz, the authors of Refs.~\cite{Rekstad1990, Sheline1995, Huseby1997} observed a $K$ dependence of the $\gamma$ decay to low-lying nuclear levels contradicting the hypothesis that the neutron-resonance states are completely mixed with respect to $K$.
In other works~\cite{Hansen1991, Barrett1992}, the same experimental observations were argued to be in agreement with the statistical model.
Still, the impact of the $K$ quantum number in highly excited deformed nuclei is not yet fully understood.

To date, no experimental data exist
for width fluctuations below neutron separation thresholds in deformed nuclei.
This region is particularly interesting
because it contains the onset of the quasicontinuum region.
There, the nuclear spectra transition
from a few individual states at low excitation energies
to an ensemble of states at high excitation energies. 
The latter are assumed to be well described by the ansatz
of the Hauser-Feshbach statistical model~\cite{HauserFeshbach1952}.
Photon-scattering reactions are frequently used to extract statistical nuclear information below particle thresholds 
including photon strength functions (PSFs)~\cite{Goriely2019db} and other \(\gamma\)-decay properties,
that are essential input to  statistical model codes~\cite{Koning2023, Kawano2016, Herman2007}.

In this Letter,
we present a quantitative study of partial width fluctuations of particle-bound $1^-$ states in the energy region primarily
associated with the concentration of \(E1\) strength called the 
pygmy dipole resonance (PDR)~\cite{Savran2013, Bracco2019, Lanza2023}.
It is the aim of this work to investigate whether the decays of $1^-$ states in this energy region can be treated as chaotic or if a certain degree of $K$ dependence is present in the studied case of $^{150}$Nd.
For this purpose, a new method was developed to study fluctuations in partial transition widths. 
It relates 
the average branching ratio of internal $\gamma$ decay transitions
to the degree of freedom $\nu$ of the \(\chi^2\) distribution.

\textit{Experiment and results---}Nuclear resonance fluorescence (NRF) experiments~\cite{Metzger1959,Zilges2022}
were performed at the Triangle Universities Nuclear Laboratory (TUNL)
with the High Intensity $\gamma$-ray Source (\higs)~\cite{Weller2009}.
The facility provides quasimonochromatic photon beams
from laser-Compton backscattering (LCB) in the \unit{\MeV} range.
The machine was operated in a new high-resolution mode. 
It reduces the spectral bandwidth of the photon beam to about \qty{1.8}{\percent}
by optimizing the electron beam parameters and using narrower apertures for the Compton-scattering geometry 
at the expense of photon-flux intensity.
The linearly polarized, quasimonochromatic \unit{\MeV}-ranged photon beams
were scattered off a target composed of \qty{11.583(1)}{\gram} of \ce{Nd_2 O_3}
enriched to \qty{93.60(2)}{\percent} of \Nd.
The target was mounted in the \gthree\ setup for \(\gamma\)-ray spectroscopy~\cite{Loeher2013}. 
It was equipped with four high-purity germanium (HPGe) detectors
placed at polar angles $\vartheta$ and azimuthal angles $\varphi$
with respect to the beam direction and its horizontal polarization plane of
$(\vartheta, \varphi) \in \{(\ang{90}, \ang{90}), (\ang{90}, \ang{180}), (\ang{135}, \ang{45}), (\ang{135}, \ang{315})\}$. 
Four lanthanum bromide (\LaBr) detectors were placed at
$(\vartheta, \varphi) \in \{(\ang{90}, \ang{0}), (\ang{90}, \ang{270}), (\ang{135}, \ang{135}), (\ang{135}, \ang{225})\}$.
A further HPGe “zero degree” detector was regularly inserted into the attenuated photon beam
to briefly measure the spectral distribution for each photon beam energy setting.

\begin{figure}[t!]
	\includegraphics{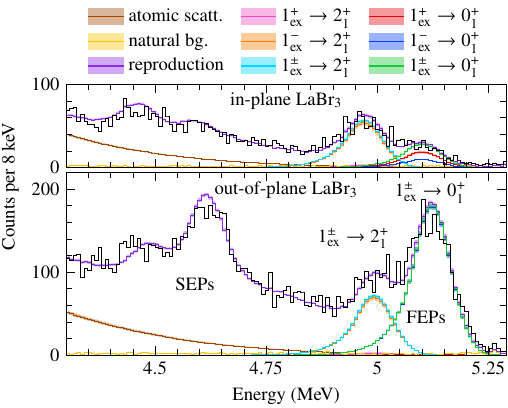}\caption{\label{fig:deconvolution}Detector response correction and decomposition of \Nd\ spectra
		for two \LaBr\ detectors.
		The fit-based reproduction is compared to the original spectrum.
		In addition, the fitted individual contributions to each spectrum are shown.
The energy ranges of full-energy peaks (FEPs) and
		single-escape peaks (SEPs) are indicated.
	}
\end{figure}

The nuclide of interest, \Nd, is close to the shape 
transitional point from spherical to prolate-deformed nuclei~\cite{Kruecken2002}.
The energy of its \(2^+_1\) state is comparatively low at \qty{130}{\keV}.
The separation of $\gamma$ transitions
from excited states to the $0^+_1$ and $2^+_1$ states of the ground band was made possible by the recently developed high-resolution mode of \higs.
The spectral distribution of the LCB beam exhibits spatial variations and, thus, can be influenced by the collimation aperture~\cite{Sun2009}. Therefore, 
the collimator radius was reduced from \qty{9.525}{\mm} to \qty{5}{\mm}
as the photon beam energy was increased from \qty{3}{\MeV} to \qty{7}{\MeV}. 
This guaranteed a beam resolution of less than \qty{130}{\keV}.
The high resolution of the LCB beam significantly reduces the overlap between the two humps containing transitions
from excited states to the \(0^+_1\) state and to the \(2^+_1\) state,
respectively (see Fig.~\ref{fig:deconvolution}).
It enables us to separate decays from excited states
to the $0^+_1$ ground state and the $2^+_1$ state
in integral spectroscopy~\cite{Isaak2019, Isaak2021},
i.e., without resolving individual transitions at all beam energies.

All spectra contain a significant signal from elastic photon scattering at the energy of the incident beam. 
It consists of contributions from excited $1^+$ and $1^-$ states,
each with its own distinct angular distribution~\cite{Krane1973,Hamilton}
with respect to \higs's linearly-polarized photon beam~\cite{Pietralla2002}. 
Excitations of $J>1$ states are  negligible in NRF reactions due to the minimal angular-momentum transfer of real photons~\cite{Zilges2022}.
Indeed, the observed angular intensity distributions agree with pure dipole radiation.
A second hump, shifted to a lower energy by the excitation energy of the $2^+_1$ state of $E(2^+_1) \approx \qty{130}{\keV}$,
corresponds to decays of excited states to the $2^+_1$ state.
A fit-based method was applied to decompose the spectrum into its individual components.
This was combined with the correction of the detector response
into a single optimization procedure per energy setting,
taking into account data from all detectors.
It was assumed that the fragmentation of states in the studied energy range
is sufficiently high, such that the excitation strength is distributed approximately uniformly
across the energy range defined by the spectral beam profile.
The latter was  measured for each photon beam energy setting
using the zero-degree detector. 
The detector responses were measured, simulated numerically, and subsequently corrected for.  
The (angular-distribution specific) detector response matrix
was simulated for each detector
using \textsc{Geant4}~\cite{Agostinelli2003,Allison2006,Allison2016}.
To reproduce the observed spectrum,
the detector response is applied separately to the four 
distinct decay branches (\(1^+\to 0^+_1, 1^-\to 0^+_1, 1^+\to 2^+_1, 1^-\to 2^+_1\)).
Natural background radiation and atomic scattering of the photon beam from the target were taken into account,
although they are negligible in the vicinity of the full-energy peak.
NRF reactions on $^{12}$C and $^{16}$O target contaminants
are corrected for using their respective detector responses
and angular distributions.
The fit simultaneously minimizes the deviation between the observed spectra
and the reproduction for all detectors, taking into account the observed background. 
Fit parameters are the number of counts of the four humps.

The measured experimental average cross sections
\(\sigma^\text{exp}\)
are the sum of
nonresonant elastic scattering processes and NRF processes~\cite{Leicht1981},
\(
	\sigma^\text{exp} \approx
	\sigma^\text{coh} +
	\sigma^\text{NRF}.
\)
For low nuclear level densities (NLDs) without overlapping levels,
this equality is exact~\cite{Rullhusen1982}.
Only at higher NLDs in the continuum region of,
e.g., the isovector giant dipole resonance (IVGDR)~\cite{Bortignon1998,Harakeh2001},
an additional term
arises from the interference between
coherent elastic scattering and NRF scattering amplitudes~\cite{Kahane1983, Kleemann2025}.
In the absence of overlapping NRF resonances,
the contributions from constructive and destructive interference
cancel each other in integral spectroscopy
and the interference term vanishes.
Multiple coherent elastic scattering processes can occur:
Delbrück, Rayleigh, and Thomson scattering.
Delbrück and Rayleigh scattering can be neglected
for the present measurement~\cite{Kane1986}
because they become significant only at very forward scattering angles where no data were taken.
Only nuclear Thomson scattering contributes significantly
to the experimentally observed cross sections~\cite{Siegbahn1965}. 
Its angular distribution is indistinguishable
from \(E1\) ground-state transitions for an even-even nucleus.
Thus, averaged cross sections
for \(E1\) ground-state transitions,
as determined by the aforementioned fit procedure,
were corrected for by the energy-independent Thomson cross section 
\(\sigma^\text{Thomson} = 8\pi/3\,(Z^2\alpha\,\hbar c/Mc^2)^2 = \qty{0.115}{\milli\barn} \)  for \Nd{}.
A table with the results from the integral spectroscopy analysis
is provided in the Supplemental Material~\cite{SupplementAPS}.

\begin{figure}[t]
	\includegraphics{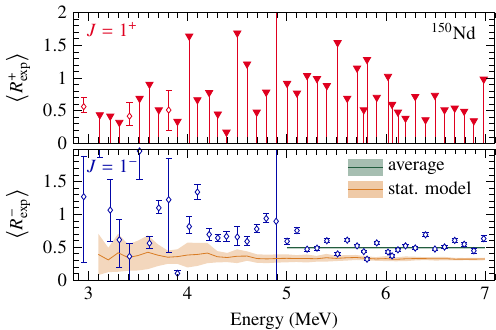}\caption{\label{fig:mean_rexp}Average branching ratios \(\braket{R_\text{exp}}\) for \Nd
		for \(1^-\) and \(1^+\) states.
		The latter are given as \(2\sigma\) upper limits
        for most energies (filled triangles), while three data points are presented with their \(1\sigma\) uncertainties (open diamonds), 
		based on the angular distribution analysis.
		For lower photon beam energies up to around \qty{4}{\MeV} to \qty{5}{\MeV},
		the fit-based method to determine the branching ratios
		is less reliable because of low NLDs
		and results in further systematic uncertainties
		that are not included in the pictured uncertainty intervals.
        A statistical model simulation based on the \textsc{Dicebox} code~\cite{Becvar1998} is also shown. See text for additional details.
	}
\end{figure}

The measured average branching ratio defined in this work
\begin{equation}\label{eq:avbr}
	\left\langle R_{\rm exp} \right\rangle =
	\frac{
        \sum_i I_{2,i}
	}{
        \sum_i I_{0,i}
	} \cdot \frac{
		E_{\gamma_0}^3
	}{
		E_{\gamma_2}^3
	} = \frac{
		\sum_i \Gamma_{0,i}
		\frac{
			\Gamma_{2,i}
		}{
			\Gamma_i
		}
	}{
		\sum_i \Gamma_{0,i} \frac{
            \Gamma_{0,i}
		}{
			\Gamma_i
		}
	} \cdot \frac{
		E_{\gamma_0}^3
	}{
		E_{\gamma_2}^3
	}
\end{equation}
depends on the energy-integrated NRF cross sections $I_{2,i}$ and $I_{0,i}$
for photoexcitation of a state \(i\) from the ground state and
subsequent decay to the  \(2^+_1\) state
and to the ground state, respectively.
It is, thus, related to the partial transition widths to the ground state \(\Gamma_{0,i}\),
to the \(2^+_1\) state \(\Gamma_{2,i}\), and the total transition width \(\Gamma_{i}\) of all levels \(i\)
populated in a given excitation-energy range, respectively.
The average \(\gamma\)-ray transition energies to the ground state and
to the \(2^+_1\) state are given by \(E_{\gamma_0}\) and \(E_{\gamma_2}\),
respectively,
and are set equal for all states in the narrow excitation-energy region
defined by the photon beam.
Experimentally, it is determined as
\begin{equation}
	\left\langle R_{\rm exp}^\pi \right\rangle =
    \frac{A_{1^\pi\to 2^+}}{A_{1^\pi\to 0^+}}
    \cdot\frac{
        \varepsilon(E_{\gamma_0}) W_{0^+ \to 1^\pi \to 0^+}
    }{
        \varepsilon(E_{\gamma_2}) W_{0^+ \to 1^\pi \to 2^+}
    }
    \cdot\frac{
		E_{\gamma_0}^3
	}{
		E_{\gamma_2}^3
	},
\end{equation}
with the total number of observed counts \(A_x\)
for the hump belonging to transition \(x\),
angular distribution coefficients \(W_x\),
and detector efficiencies \(\varepsilon(E_\gamma)\) for energy \(E_\gamma\).
The experimental \(\left\langle R_\text{exp}^\pm \right\rangle\) values are depicted 
in the lower (upper) panel for \(1^-\) (\(1^+\)) states in Fig.~\ref{fig:mean_rexp}.
The angular distributions for \(0^+\to 1^\pm\to 2^+\) cascades
are considerably less pronounced than those observed for ground-state transitions.
Consequently, the sensitivity to the parity of excited states is reduced.
Most of the decays to the \(2^+\) states are attributed to \(1^-\) states.
Since \(1^-\) states dominate the ground-state transition 
strength, only upper limits could be determined for decays of \(1^+\) states at most energies (filled triangles). 
For three measurements, the data are presented with their \(1\sigma\) uncertainty.
At lower excitation energies, up to approximately \num{4} to \qty{5}{\MeV},
the average branching ratios for \(1^-\) states vary strongly,
and their uncertainties are large.
This phenomenon can be attributed to the low NLD. 
It violates the assumption employed in the fit procedure
that the strength is distributed uniformly across the excitation-energy region. Thus, the decay properties of individual states dominate the branchings.
The situation is different for excitation energies exceeding \qty{5}{\MeV}. 
Here, higher NLDs lead to reliable results
for the average branching ratios.
In this region, the \( \left\langle R_\text{exp}^- \right\rangle \) values
do not vary strongly and converge upon a common value of about \num{0.5}.
For the following analysis, an average value of 
\(\bigl\langle\langle R_\text{exp}^-\rangle\bigr\rangle = \num{0.490(16)}\) 
was determined for energies above \qty{5}{\MeV} using a Bayesian approach
accounting for the scatter of the data points~\cite{Brooks2011,footnote_frequentist}.

\textit{Discussion---}The observed average branching ratios are discussed
with respect to the statistical Hauser-Feshbach model~\cite{HauserFeshbach1952}.
The partial transition widths \(\Gamma_{j,i}\) in Eq.~(\ref{eq:avbr}) are, thus, considered to be \(\chi^2\) distributed (normalized to unit expectation) 
and given by 
\(\Gamma_{j,i} = \bigl\langle\Gamma_j\bigr\rangle \cdot f_i\)
for a \(0^+_1 \to 1^\pm_i \to J_j^+\) cascade,
where \(f_i\) is a \(\chi^2\)-distributed random variable
normalized to its degree of freedom \(\nu\).
The total transition widths are sums of \(\chi^2\)-distributed quantities. 
They can be approximated using a Gaussian distribution represented by the random variable, \(g_i\),
if more than three partial widths contribute to 
each \(\Gamma_i\)~\cite{SupplementAPS}.
Thus, one can transform the ratio
of sums of partial and total transition widths in Eq.~(\ref{eq:avbr}) into
\begin{align}
    \label{eq:ratio_avg_ggg}
    \dfrac{
        \sum_{i} \Gamma_{0,i} \frac{\Gamma_{2,i}}{\Gamma_i}
    }{
        \sum_{i} \Gamma_{0,i} \frac{\Gamma_{0,i}}{\Gamma_i}
    }  & =
    \frac{
    \left\langle \Gamma_{2} \right\rangle
    }{
    \left\langle \Gamma_{0} \right\rangle
    }
    \dfrac{
        \sum_{i}
        f_i \cdot
        f'_i
        / g_i
    }{
        \sum_{i}
        f_i \cdot
        f_i
        / g_i
    }
    =
    \frac{
    \left\langle \Gamma_{2} \right\rangle
    }{
    \left\langle \Gamma_{0} \right\rangle
    }
    s     . 
\end{align}
The quantity is separated into a factor
that depends on the ratio of average partial transition widths
to the ground state (\(2^+_1\) state), given by
\(\left\langle \Gamma_{0} \right\rangle\) (\(\left\langle \Gamma_{2} \right\rangle\))
and the internal fluctuation ratio 
\begin{align}
 \label{eq:s_value}
 s & \equiv
    \dfrac{
        \sum_{i}
        f_i \cdot
        f'_i
        / g_i
    }{
        \sum_{i}
        f_i \cdot
        f_i
        / g_i
    }.
\end{align}
It is sensitive to the statistical distribution
of partial transition widths of excited states.
For \(\chi^2\) distributions,
the internal fluctuation ratio \(s\)
can be directly related to the degree of freedom \(\nu\)~\cite{SupplementAPS} via
\begin{equation}
    \label{eq:fluctuation_factor}
    s = \frac{\nu}{\nu + 2}
    \qquad \Leftrightarrow \qquad
    \nu = \frac{2s}{1 - s}.
\end{equation}
A detailed analysis of the convergence of \(s\)
as a function of the number of excited states and decay branches
can be found in the Supplemental Material~\cite{SupplementAPS}.
Assuming all partial widths being independently PT distributed,
i.e., \(\nu = 1\), a value of \(s=1/3\) is obtained; see Refs.~\cite{Axel1962,Bartholomew1973} for similar conclusions.
It is stressed, that \(s=1/3\) only if the ground-state decay channel is involved, while otherwise \( s = 1 \).
This follows from the multiplicativity 
of the expected value of independent random variables, which leads to the cancellation of the terms in Eq.~(\ref{eq:s_value}). 
For a detailed discussion see Ref.~\cite{SupplementAPS}.

As a result, under the assumption that all observed transitions
are accurately described by the statistical model, i.e,
given all partial transition widths follow the same \(\chi^2\) distribution,
\( \left\langle R_{\rm exp}^- \right\rangle \) is related to \(s\) via
\begin{align}
    \label{eq:Rexp_s}
    \braket{R_{\rm exp}^-} =
    s \dfrac{
        \braket{\Gamma_2}_{E1}
    }{
        \braket{\Gamma_0}_{E1}
    }
    \dfrac{
        E_{\gamma_0}^3
    }{
        E_{\gamma_2}^3
    } =
    s \dfrac{
        f_{E1}(E_{\gamma_2})
    }{
        f_{E1}(E_{\gamma_0})
    }
\end{align}
with \(f_{E1}(E_{\gamma_i}) \propto \braket{\Gamma_i}_{E1} / E_{\gamma_i}^3\)~\cite{Bartholomew1973, Goriely2019db} being the PSF for $E1$ transitions with \(\gamma\)-ray energies \(E_{\gamma_i}\).
Assuming the validity of the Brink-Axel (BA) hypothesis~\cite{BrinkPhD,Axel1962},
\(f_{E1}\) is expected to differ by less than \qty{3}{\percent} within the \qty{130}{\keV} difference for 
decays to the \(0^+_1\) state and to the \(2^+_1\) state of \Nd in the region of interest from \num{5} to \qty{7}{\MeV}.
Consequently, in the case of a typical deformed rare-earth nucleus like \Nd
with a low \(2^+_1\) excitation energy,
Eq.~(\ref{eq:Rexp_s}) can be approximated by \(\left\langle R_{\rm exp}^- \right\rangle \approx s \).

To validate the formalism derived above,
a statistical model simulation was performed
using a version of the Monte-Carlo method-based \textsc{Dicebox} code~\cite{Becvar1998}
adapted for the simulation of \gray{} cascades in NRF reactions.
Starting from the 12 lowest-lying experimentally known levels of \Nd~\cite{nds150},
an artificial nucleus is generated (a so-called realization)
with a random level scheme with partial decay widths for each level following PT fluctuations.
The average decay behavior follows the statistical properties
defined by the input NLD and PSF.
In order to calculate the NLD,
the back-shifted Fermi gas model is employed, with parameters
\(E1 = \qty{-0.516}{\MeV}\) and \(a = \qty{16.275}{\per\MeV}\) taken from Ref.~\cite{vonEgidy2009}.
The input PSFs are composed of the IVGDR
(standard Lorentzians, parametrized according to Ref.~\cite{Carlos1971})
for \(E1\) contributions, and the 
scissors and spin-flip resonances parametrized by the simple modified Lorentzian model~\cite{Goriely2019} to account for \(M1\) transitions
and negligible \(E2\) contributions from single-particle contributions.
For each photon beam energy setting of the experiment,
the population of excited states via the NRF reaction is simulated
in accordance with the actual spectral distribution
of the photon beam.
The subsequent decay of these states via \gray{} cascades
is simulated and analyzed in the same way
as the experimental data.
Because of the low energy of the \(2^+_1\) state,
the results of the statistical model simulation
for the average branching ratio
are not sensitive to the PSF and NLD model parameters. 
Similar results are obtained for any reasonable combination of values.

The entire process is repeated for a total of 30 different realizations.
For each photon beam energy setting,
\( \left\langle R_\text{sim}^- \right\rangle \) is extracted from the simulations
in a manner analogous to the experimental results of this work.
A comparison of the simulated 
(orange solid line and uncertainty band) and the experimental results is presented in the lower panel of Fig.~\ref{fig:mean_rexp}.
For both, experiment and simulation,
the average $E1$ branching ratio
stays approximately constant between \num{5} and \qty{7}{\MeV}.
The simulation yields \(\left\langle R_\text{sim}^- \right\rangle \approx \num{0.31}\),
which is in excellent agreement with the internal fluctuation ratio of \(s = 1/3\)
required for PT-distributed partial transition widths.
Thus, it follows that the derivation of \(s\)
and the approximations that lead to \(\left\langle R_{\rm exp}^- \right\rangle \approx s \) are well justified.

For the experimental result of \(\bigl\langle\langle R_\text{exp}^-\rangle\bigr\rangle = \num{0.490(16)}\) above \qty{5}{\MeV},
a value of \(\nu=\num{1.93(12)}\) is obtained using Eq.~(\ref{eq:fluctuation_factor}).
This value is considerably larger than \(\nu=1\). 
It disagrees with a pure PT distribution by 7 standard deviations.
Photon strength functions are usually parametrized with (modified) Lorentzian functions~\cite{Goriely2019db} and, thus,
typically increase monotonically with $\gamma$-ray energy, i.e. \(f_{E1}(E_{\gamma_2}) / f_{E1}(E_{\gamma_0}) < 1\),
leading to even larger values for \(s\) and, consequently, \(\nu\).
It can, therefore, be concluded that
the assumptions of the statistical model are inconsistent
with the experimental observations.

Besides a modified statistical distribution of partial transition widths,
namely a modified degree of freedom \(\nu\)
of the \(\chi^2\) distribution,
other explanations may account for the observed deviations:
The assumption of large NLDs (\(> 100\) levels per \unit{\MeV}) could be violated,
or strong nonstatistical decays could
skew the resulting distribution of partial transition widths
in a way that cannot be described by the \(\chi^2\) distribution.

A potential nonstatistical contribution to the observed decay behavior may arise from the decay pattern of nuclear states in a well-deformed nucleus, according to the Alaga rule~\cite{Alaga1955}.
The decay intensity for transitions to the ground-state rotational band depends on the excited states' quantum number $K$, where $K$ is the projection of the total angular momentum onto the symmetry axis of the nucleus.
For states with quantum number $K=0$, the expected branching ratio is \( \Gamma_2 / \Gamma_0 \cdot (E_{\gamma_0} / E_{\gamma_2})^3 = 2.0 \), while for $K=1$ states it is \( 0.5 \).
The experimental branching ratios below \qty{4}{\MeV} for \Nd{}
were found to be in reasonable agreement with the Alaga rules~\cite{Pitz1990}. 
Furthermore, the three data points below \qty{4}{\MeV} in Fig.~\ref{fig:mean_rexp},
which show significant contributions from $1^+$ states (open diamonds in the upper panel), 
are in agreement with the Alaga rule for quantum number $K=1$ which is expected for the low-energy $M1$ strength.
Previous studies on deformed nuclei
exhibit a preference for \(K=0\)
for low-energy \(1^-\) states below \qty{4}{\MeV}~\cite{Savran2005}.
To estimate the strength from nonstatistically decaying states, 
\begin{align}
    \braket{R_{\rm exp}^-} & =
    s \dfrac{
        f_{E1}(E_{\gamma_2})
    }{
        f_{E1}(E_{\gamma_0})
    }
    C_{\rm stat} + 2 C_{K=0} + 0.5 C_{K=1}
\end{align}
can be decomposed into contributions from excited $1^-$ states that either decay statistically (\(C_{\rm stat}\)) or exhibit  pure \(K=0\) quantum number (\(C_{K=0}\)), or \(K=1\)  (\(C_{K=1}\)) with the identity \(C_{\rm stat} + C_{K=0} + C_{K=1} = 1\).
A detailed derivation is provided in the Supplemental Material~\cite{SupplementAPS}.

Assuming that only $K=0$ states contribute to the nonstatistical part, i.e. \(C_{K=1} = 0\), using \(\braket{\braket{R_{\rm exp}^-}} = \num{0.490(16)}\) and \(s = 1/3\), one finds that even a small contribution of
\(C_{K=0} = \qty{9.4(10)}{\percent}\) of the total ground-state decay cross section from \(K=0\) states is sufficient to explain the experimental observations.
It is important to note that this value marks a lower limit of the nonstatistical contribution.
An upper limit of \(C_{K=1} = \qty{94(10)}{\percent}\) is obtained, if solely \(K=1\) states would contribute, implying \(C_{K=0} = 0\).

These findings strongly support previous evidence that the  decay properties of
photoexcited $1^-$ states in the quasicontinuum regime do not follow the expectations from the
statistical model~\cite{Angell2012, Isaak2013, Isaak2019, Papst2020, Sieja2023}.
Moreover, the fluctuations of partial transition widths of $1^-$ states in the region of the PDR 
to the ground-state band do not follow the PT distribution. Therefore, these $E1$ transitions cannot be treated as a 
representation of an entirely chaotic ensemble, even in a nucleus with high NLD such as $^{150}$Nd. Contributions from nuclear structure properties of deformed nuclei related to the $K$ quantum number may play a significant role.
It is noted, that fluctuation properties of partial widths,
especially for the ground-state decay channel, need to be treated correctly in NRF reactions.
If average branching ratios \( \left\langle R_{\rm exp}^\pm \right\rangle \)
are related to the ratio of PSFs,
the internal fluctuation ratio \( s \) must be taken into account.
This is especially important
if values for PSFs are extracted via the ratio or shape method~\cite{Wiedeking2012, Wiedeking2021}.

The photoabsorption cross section of \Nd{}
reveals a pronounced double-humped structure in the region
of its isovector giant dipole resonance (IVGDR)~\cite{Carlos1971}.
In more recent experiments,
the observed splitting is less pronounced~\cite{Donaldson2018}.
This splitting of the IVGDR is widely regarded as a signature of nuclear deformation and
is attributed to axial deformation, which enables proton-neutron vibrations to occur
parallel ($K=0$) and perpendicular ($K=1$) relative to the nuclear symmetry axis
with different associated frequencies~\cite{Danos1958, Okamoto1958}.
Recent observations of the $\gamma$ decay from the IVGDR in the well-deformed nucleus \Sm{} revealed a distinct variation
of \(\braket{R_{\rm exp}^-}\) across the two humps
identified as the $K=0$ and $K=1$ IVGDR components~\cite{Kleemann2025}.
In hydrodynamical models, the PDR is usually interpreted as a collective oscillation of a neutron skin relative
to an inert isospin-saturated core.
Accordingly, a splitting of the PDR strength in deformed nuclei, analogous to the behavior observed for the IVGDR,
might be anticipated~\cite{Lanza2023}.
It is noteworthy that the effect of the deformation
on the dipole response in the region of the PDR is not conclusive at all~\cite{Lanza2023}.
The present study finds a rather constant value of \(\braket{\braket{R_\text{exp}^-}} = \num{0.490(16)}\) in the energy
region from \num{5} to \qty{7}{MeV} for \Nd{}, with no evidence of an abrupt or even gradual transition
between regions corresponding to different $K$ quantum numbers, as observed in the IVGDR of \Sm{}~\cite{Kleemann2025}.
Consequently, the data presented here for the energy region below \qty{7}{MeV} provide no indication of
a pronounced splitting of the $E1$ strength into $K$ components in the PDR region of \Nd{}.

\textit{Summary---}
For the first time,
the statistical distribution of partial transition widths
below particle separation thresholds could be probed using NRF experiments
utilizing the new high-resolution \higs operating mode.
An internal fluctuation ratio of \(s = \num{0.490(16)}\) was observed,
which differs significantly from the value 
\(s = 1/3\) predicted for PT-distributed partial transition widths 
and confirmed by statistical model simulations.
Assuming \(\chi^2\)-distributed partial transition widths,
the observed internal fluctuation ratio is associated
with a degree of freedom of \(\nu = \num{1.93(12)}\) in clear disagreement with the PT distribution.
Contributions as small as \qty{9.4(10)}{\percent}
and up to \qty{94(10)}{\percent}
from nonstatistical $\gamma$ decays
were shown to be capable of significantly altering 
the average $\gamma$-decay behavior in agreement 
with the data on $^{150}$Nd.
They demonstrate that an entirely statistical description 
of $E1$ decays into the ground-state band from an excitation energy region which exhibits a NLD for $1^-$ states in the order of $10^3$ levels per MeV is inaccurate. 
Whether this discrepancy originates in the survival of some states with pure $K$ quantum numbers or in an expression of a specific average $K$-mixing remains an open question. 
Moreover, good $K$-quantum numbers are strictly valid only in well-deformed axially symmetrical nuclei.
In the case of spherical nuclei, the situation can be quite different.
Since the presented new approach is applicable to any stable isotope on the nuclear chart,
systematic studies for several spherical and deformed nuclei are underway to address this question.

\begin{acknowledgments}
    We would like to thank A. Leviatan for useful comments and discussions.
    We thank
    \mbox{S.~Bassauer},
    \mbox{M.~Berger},
    \mbox{P.~C.~Ries},
    \mbox{E.~Hoemann},
    \mbox{N.~Kelly},
    \mbox{R.~Kern},
    \mbox{Krishichayan},
    \mbox{D.~R.~Little},
    \mbox{M.~Müscher},
    \mbox{J.~Sinclair}, and
    \mbox{J.~Wiederhold}
    for help with the data taking.
    We thank the \higs\ accelerator crew for providing excellent photon beams for our experiment.
    The \Nd{} target was supplied by the Isotope Program
    within the Office of Nuclear Physics in the U.S. Department of Energy's Office of Science.
    This work has been funded
    by the Deutsche Forschungsgemeinschaft (DFG, German Research Foundation)---Project-ID 279384907---SFB~1245,
    Project-ID 460150577---ZI 510/10-2,
    and Project-ID 499256822---GRK~2891 “Nuclear Photonics”,
    by the State of Hesse under the grant “Nuclear Photonics” within the LOEWE program (LOEWE/2/11/519/03/04.001(0008)/62),
    by the BMBF under Grant No. 05P21RDEN9,
    and by the U.S. Department of Energy, Office of Science, Office of Nuclear Physics,
    under Grant No. DE-SC0023010, No. DE-FG02-97ER41041 (UNC), and No. DE-FG02-97ER41033 (Duke-TUNL).
    M. Scheck acknowledges financial support by the UK-STFC (Grant No. ST/P005101/1).
\end{acknowledgments}

\textit{Data availability---}Data to support Fig.~\ref{fig:mean_rexp} are available within the Letter
and the corresponding Supplemental Material~\cite{SupplementAPS}.
All data corresponding to findings in this Letter
are openly available at the TUdatalib repository of Technische Universität Darmstadt~\cite{Papst2024dataset}
and further details are available in Ref.~\cite{Papst2024thesis}

\bibliography{bibliography}

\begin{thebibliography}{91}
\bibitem [{foo()}]{footnote_frequentist}%
  \BibitemOpen
  \href@noop {} {\bibinfo {title} {{In a frequentist approach, the uncertainty
  of \( \bigl\langle\langle R_\text{exp}^-\rangle\bigr\rangle \) comprises a
  "statistical" component of $0.006$ and an additional "systematic" component
  of $0.010$, which accounts for the observed scatter. The total uncertainty of
  $0.016$ is the sum of these two components.}}}\BibitemShut {Stop}%
\end{thebibliography}

\end{document}